\documentclass{natureprintstyle2}
\usepackage{graphicx,amsmath}
\usepackage{astjnlabbrev-nature} 

\usepackage{amssymb}
\title{Turning off the lights: Supernova SN1987A 30 years on}

\author{Richard de Grijs$^{1,2,3}$}
\begin{document}

\maketitle

\begin{affiliations}
 \item Kavli Institute for Astronomy and Astrophysics, Peking
   University, Yi He Yuan Lu 5, Hai Dian District, Beijing 100871,
   China; e-mail: grijs@pku.edu.cn\\
 \item Department of Astronomy, Peking University, Yi He Yuan Lu 5,
   Hai Dian District, Beijing 100871, China\\
 \item Discipline Scientist, International Space Science
   Institute--Beijing\\
\end{affiliations}

\begin{abstract} 
Decades-long repeat observations of supernova SN1987A offer us unique,
real-time insights into the violent death of a massive star and its
long-term environmental effects, until its eventual switch-off.
\end{abstract}

On 23 February 1987, supernova SN1987A---the explosive last gasps of a
dying massive star---suddenly illuminated the Large Magellanic
Cloud. Such bright stellar explosions occur only once or twice per
century in large spiral galaxies like our Milky Way and much less
frequently in Magellanic-type ‘dwarf’ galaxies. At a distance of
163,000 light-years\cite{dg14}, SN1987A was the closest recorded
supernova since Kepler's Supernova in 1604 and Cassiopeia A in the
late-17$^{\rm th}$ Century, both of which occurred in the Milky
Way. Fortunately, we now have access to cutting-edge astronomical
facilities, thus allowing us to monitor the explosion and subsequent
evolution of the entire SN1987A system in real time.

In a recent article in {\it The Astrophysical Journal Letters},
Fransson {\it et al.}\cite{fr15} present a 20-year timeline of changes
in the appearance of the supernova's complex system of rings. Two of
these rings are most likely caused by mass outflows from its red
supergiant progenitor star up to 20,000 years before the
explosion\cite{cr91}. The brighter but smaller inner ring, on the
other hand, was generated by interactions of the progenitor's swept-up
stellar wind with the ambient gas. Hubble Space Telescope observations
have enabled us to witness the appearance and subsequent disappearance
of `hot spots' along this latter circumstellar ring. These are likely
caused by interactions of the densest gas clumps in the circumstellar
gas with the supernova's outward-propagating blast wave, thus leading
to the appearance of a ring-like shape. The expanding supernova debris
left behind after the blast wave passed by is decelerated by a
`reverse shock', which is due to electrons cascading down to lower
energy levels following collisional excitation of neutral hydrogen
atoms, triggered when the debris crosses the shock front.

Until approximately 2009, the observational timeline of Fransson {\it
  et al.}\cite{fr15} shows an exponentially increasing contribution to
the hot spots from shocked emission and outward acceleration of these
clumps of up to 700--1000 km s$^{-1}$. New hot spots and faint,
diffuse emission have since appeared outside of the now-fading inner
ring. These new hot spots could have been triggered by the expanding
blast wave if the density of the interstellar clumps was sufficiently
high for the pressure from the reverse shock to cause cooling by
radiating away heat. This, in turn, would cause the clumps to collapse
into an even smaller volume and emit optically visible
radiation\cite{pun02}. Ionization of the original stellar-wind ejecta
caused by X-ray emission originating from lower-density material
behind the expanding shock wave may be responsible for the diffuse
emission.

The authors\cite{fr15} interpret the decrease in optical emission in
the inner ring since 2013 as evidence of its dissolution. They deduce
that the most likely reason for this decrease is that the area
affected by shocks generating bright optical radiation is getting
smaller. Alternatively, the expected steeply outward-increasing radial
density distribution of the stellar-wind material\cite{dew12,fr13}
implies a corresponding increasing density threshold for clumps to
sustain such radiative shocks. Either of these scenarios will lead to
instabilities in the hot spots, conduction by the ambient hot gas and,
consequently, rapid dissolution of the densest
clumps\cite{pun02}. Fransson {\it et al.}\cite{fr15} predict that the
inner ring will likely be fully destroyed within the next 10 years.

\begin{figure}
\includegraphics[width=\columnwidth]{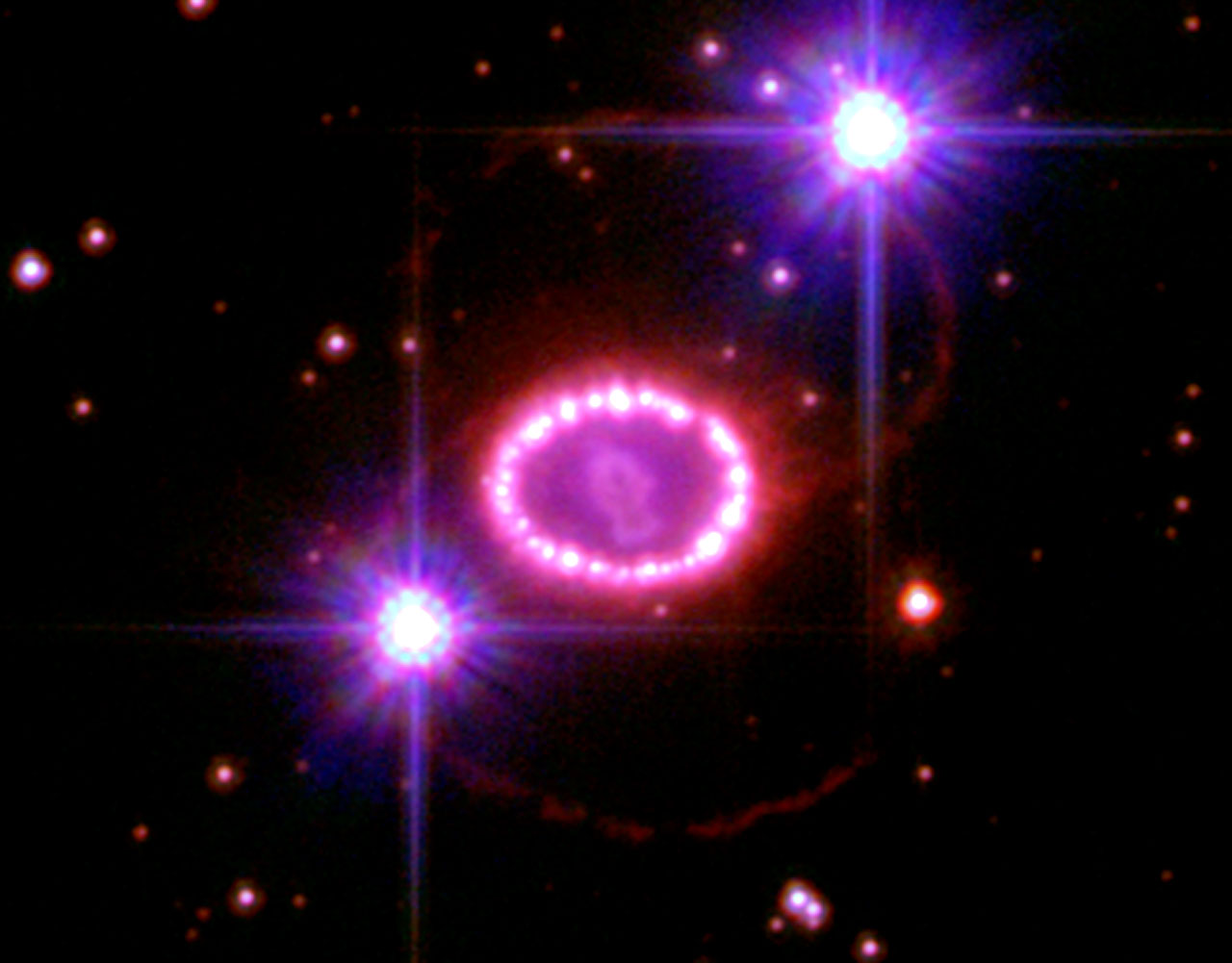}
\begin{center}
\caption{Optical Hubble Space Telescope image of SN1987A taken in
  December 2006 with the Advanced Camera for Surveys. (Credit: NASA,
  ESA, and R. Kirshner, Harvard--Smithsonian Center for Astrophysics)}
\end{center}
\label{F1}
\end{figure}

SN1987A thus provides unprecedented insights into the physical
properties of its massive progenitor from a perspective that is
unavailable elsewhere. The system's impact goes well beyond this
immediate application. We can, in principle, also use the maximum
observed angular size of the ring, combined with the speed of light
and accurate delay-time measurements, to independently obtain a
reliable geometric distance to the supernova. However, since the ring
is resolved, we need to consider whether the emission used to measure
delay times at a variety of wavelengths actually originates from the
same region(s) in the ring. For instance, the ring size measured using
optical emission from doubly ionized oxygen atoms\cite{pl95} is
significantly different from that traced by ultraviolet
lines\cite{pa91}. It has been suggested that the latter originate from
the ring's inner edge, whereas the optical lines come from its main
body. In this case, and using the proper geometry, including a finite
ring thickness, the ultraviolet light curve could result in an
underestimate of the light-travel time across the optical ring
diameter of up to 7\% and, thus, a similar underestimate of the
distance\cite{gu98}.

We also need to consider potential errors caused by a
misinterpretation of the underlying physics. Most importantly, it is
often assumed that the fluorescent, scattered emission from the
interstellar gas commences as soon as the supernova's energetic
photons hit a gas cloud. It is possible, though, that there is a
slight delay in the onset of fluorescent emission as the gas first
recombines from highly ionized states, for instance. Neglecting this
step will lead to ring-size and, hence, distance overestimates.

SN1987A will likely become more X-ray and less hot-spot dominated as
the system continues to evolve\cite{fr15}. The high-resolution Hubble
Space Telescope observations of Fransson {\it et al.}\cite{fr15} and
their spectroscopic analysis will allow us to settle the system's full
evolution unequivocally and provide the tightest geometric constraints
yet on the use of resolved supernovae as distance tracers. As the
important first rung of the extragalactic distance ladder, the
importance of reducing the systematic uncertainties in the distance to
the Large Magellanic Cloud cannot be overstated\cite{dg14}.

\end{document}